          [2001/04/25 1.1 (PWD)]
\documentclass[a4paper,twocolumn]{esapub2005} % European paper
\usepackage{times}
\usepackage{natbib}
\usepackage{graphicx}

\title{Asteroseismology of exoplanets-host stars}
\author{Sylvie Vauclair}
\affil{Laboratoire d'Astrophysique de Toulouse Tarbes; Observatoire Midi-Pyr\'en\'ees; Universit\'e Paul Sabatier, Toulouse, France}

\begin{document}


\maketitle

\begin{abstract}
Studying the internal structure of exoplanets-host stars compared to that of similar stars without detected planets is particularly important for the understanding of planetary formation. 
The observed overmetallicity of stars around which planets have been detected may be a hint in that respect. In this framework, asteroseismic studies represent an excellent tool to 
determine the structural differences between stars with and without detected planets. After a general discussion on this subject, I present the special cases of three different stars: $\mu$ Arae which has been observed with the HARPS spectrograph in June 2004, $\iota$ Horologii, that we have studied in detail and will be observed with HARPS in November 2006, and finally HD 52265, one of the main targets of the COROT mission, an exoplanets-host star which will be observed with the COROT satellite during five consecutive months.
\end{abstract}

\section{Introduction}

Since the first discovery of a planet orbiting around Peg 51 (Mayor \& Queloz \cite{mayor95}), nearly 200 exoplanets have been detected. Due to the bias of the detection techniques (radial velocity or transit methods), the observed planetary systems are different from the Solar System: only planets orbiting close to the central star can lead to observable effects. Most of them are “hot Jupiters”, i.e. Jupiter-like planets at distances of order or less than one astronomical unit. 

The central stars of these planetary systems appear to be overmetallic compared with the Sun, at least in their atmospheres. Their average metallicity is $\sim$ 0.2 dex larger than solar while it is about solar for stars which have no detected planets (Santos et al. \cite{santos03} and \cite{santos05}, Gonzalez \cite{gonzalez03}, Fischer and Valenti \cite{fischer05}). Two scenarios have been proposed to explain these high metallicities (see Bazot \& Vauclair \cite{bazot04}). In the first scenario, they are the result of a high initial metal content in the proto-stellar gas while in the second scenario, the overmetallicity is due to the accretion of hydrogen-poor matter during planetary formation. 

The initial overmetallicity scenario seems more probable than the accretion scenario. The first argument which was given against accretion was that the observed overmetallicity does not vary with the stellar mass while the masses of the outer convective zones may strongly differ. This argument does not hold however, due to extra convection induced by the inverse $\mu$-gradient (Vauclair \cite{vauclair04}). On the other hand, the huge mass which should be accreted while the star is already on the main sequence (about 100 earths) does not seem realistic. 
Another argument generally given in favor of the primordial overmetallicity is related to planet formation, as the metallic overabundance should help planet condensation. 

One should however keep in mind that the stars for which planets have been detected host giant planets orbiting at small distances, which means that these planets have migrated from their formation site, far from the star, to their present location. On the other hand, the stars for which no planets have been detected may host planets as well as the other ones, if only these planets have not migrated (this is the case of the solar system). As a consequence, the difference between the so-called exoplanets-host stars and the other ones is not a question of planetary formation, but a question of migration: why is it that in some cases the planets move toward the star and not in the other cases?

This question has no answer yet. In this framework, asteroseismic studies represent an excellent tool to help 
determining the structural differences between stars with and without detected planets. 
Bazot \& Vauclair \cite{bazot04} studied the internal differences of stellar models computed with either the overmetallic or the accretion assumption, but 
iterated so as to present exactly the same observable parameters (T$_{eff}$, $\frac{L}{L_{\odot}}$, Z, log g): these models can account for the same observed star, but their interiors are different. 
Moreover, they may have different masses while their observed parameters are the same. It is also possible that among two models representing the same observed star, one has a convective core while the other does not. 

In the following sections, I first give some generalities about the modelisation procedures, and then I present the case of three particular stars: HD 160691 (alias $\mu$ Arae) which has been observed for seismology with the HARPS spectrograph in June 2004, HD 17051 (alias $\iota$ Horologii), that we have studied in detail and will be observed with HARPS in November 2006, and finally HD 52265, one of the main targets of the COROT mission, an exoplanets-host star which will be observed during five consecutive months.

\section{Modelisation }

In all the computations and comparisons presented below, the models were computed using the TGEC (Toulouse-Geneva stellar evolution code), with the OPAL equation of state and opacities (Rogers \& Nayfonov \cite{rogers02}, Iglesias \& Rogers \cite{iglesias96}) and the NACRE nuclear reaction rates (Angulo et al.\cite{angulo99}). In all models microscopic diffusion was included using the Paquette prescription (Paquette et al. \cite{paquette86}, Richard et al. \cite{richard04}). The treatment of convection was done in the framework of the mixing length theory and the mixing length parameter was adjusted as in the solar models ($\alpha = 1.8$), except in some specific cases which are specially discussed. Different kinds of models were computed, according to the initial assumptions: overmetallic models with two different initial helium values, and accretion models. 

For the computations of overmetallic models, the helium value is crucial as differences in helium may lead to completely different evolutionary tracks. For the computations of the accretion models, the way accretion occurred, as well as the composition of the accreted matter is also a subject of debate. In the models, the most simple assumption was used: instantaneous accretion of matter with solar composition for metals and no light elements, at the beginning of the main-sequence. Neither extra-mixing nor overshoot were included.

Adiabatic oscillation frequencies were computed using the PULSE code adapted from Brassard \cite{brassard92}. For each evolutionary track, many models were computed inside the observed spectroscopic boxes (color-magnitude or log g–log T$_{eff}$ diagrams, as described below). The most useful combinations of the oscillation frequencies (large and small separations, echelle diagrams) were computed for comparisons with already available, or future observational data. 

\section{The example of $\mu$ Arae }

The exoplanets-host star $\mu$ Arae is a G5V star with a visual magnitude V=5.1, an Hipparcos parallax $\pi$=65.5$\pm$0.8 mas, which gives a distance to the Sun of 15.3 pc and a luminosity of $\log L/L_{\odot}=$0.28$\pm$0.012. Spectroscopic observations by various authors gave five different effective temperatures and metallicities (see references in Bazot et al. \cite{bazot05}). The HARPS observations allowed to identify 43 oscillation modes of degrees l=0 to l=3 (Bouchy et al. \cite{bouchy05}). From the analysis of the frequencies and comparison with models, the values T$_{eff}$=5813$\pm$40 K and [Fe/H]=0.32$\pm$0.05 dex were derived: these values, which lie inside the spectroscopic boxes, were obtained with a much better precision than spectroscopy.

For each evolutionary track, many models were computed inside the observed spectroscopic boxes in the HR diagram, but only those which could reproduce the observed echelle diagram were kept for subsequent tests (Figures 1 and 2). For these models the large separation $\Delta \nu_{l} = \nu_{n+1, l} - \nu_{n, l}$ is exactly 90 $\mu$Hz: a difference of only 0.5 $\mu$Hz completely destroys the fit with the observations. 

\begin{figure}
\centering
\includegraphics[width=0.8\linewidth]{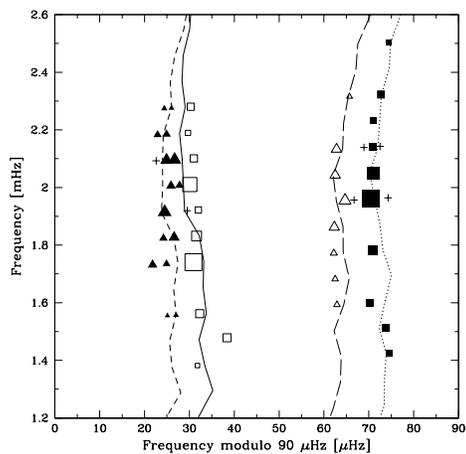}
\caption{ Echelle diagram for a model of the star $\mu$ Arae, computed with the assumption of primordial overmetallicity. The ordinate represents the frequencies of the modes and the absissa the same frequencies modulo the large separation, here 90 $\mu$Hz. The lines represent the computations, respectively, from left to right : l=2, 0, 3 and 1 and the symbols represent the observations (after Bazot et al. \cite{bazot05}).}
\label{fig1}
\end{figure}

\begin{figure}
\centering
\includegraphics[width=0.8\linewidth]{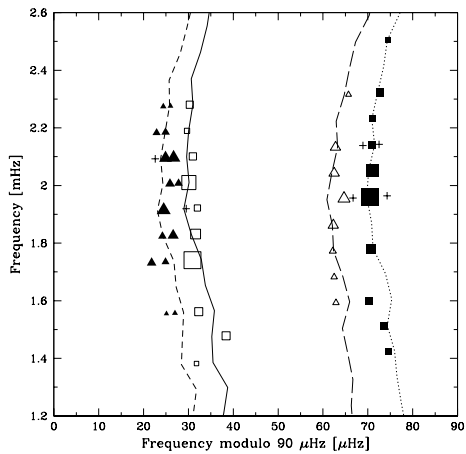}
\caption{Same as Figure 1 for a model computed with accretion.}
\label{fig2}
\end{figure}

We can see a clear difference between the overmetallic and the accretion cases for the lines $l=0$ and $l=2$: in the overmetallic case they come closer at large frequencies and even cross around $\nu = 2.7$ mHz, which does not happen in the accretion case. This behavior clearly appears in the representations of the small separations (see specific figures in Bazot et al. \cite{bazot05}). I will come back on this important effect below, for another star. For $\mu$ Arae it was not yet possible to decide which scenario was the best one, in spite of the very good data obtained, but we are still working on it.

\section{Modelisation of the particular star $\iota$ Horologii}

Among exoplanets-host stars, $\iota$ Hor is a special case for several reasons. Three different groups have given different stellar 
parameters for this star (table 1). Meanwhile, Santos et al. \cite{santos04} suggest a mass of 1.32 M$_{\odot}$ while Fischer and Valenti \cite{fischer05} give 1.17 M$_{\odot}$.

Some authors (Grenon \cite{grenon00}, Chereul and Grenon \cite{chereul00}, Kalas and Delorn \cite{kalas06}) pointed out that this star has the same kinematical characteristics as the Hyades: as well as many other stars observed in the whole sky, it belongs to the ``hyades stream'': they all point towards the convergent. This may be due to dynamical effects in the Galaxy, but it is also possible that the star has been formed with the cluster and moved out due to evaporation. In this case, it would mean that its overmetallicity is primordial. 

\begin{table}
\caption{Observed parameters for $\iota$ Hor (after Laymand \& Vauclair \cite{laymand06}); a) Gonzalez et al. 2001; b) Santos et al. 2004; c) Fischer and Valentini 2005}
\label{tab1}
\begin{center}
\begin{tabular}{cccc} \hline
\hline
T$_{\mbox{eff}} (K)$ & log g  & [Fe/H] & Ref. \cr
 \hline
6136$\pm$34 & 4.47$\pm$0.05 & 0.19$\pm$0.03 & a\cr
6252$\pm$53 & 4.61$\pm$0.16 & 0.26$\pm$0.06 & b\cr 
6097$\pm$44 & 4.34$\pm$0.06 & 0.11$\pm$0.03 & c\cr
\hline
\end{tabular}
\end{center}
\end{table}

Laymand \& Vauclair \cite{laymand06} computed evolutionary tracks and models lying inside the error boxes as given by the observers. In their paper, the oscillation frequencies for several characteristic models are presented and some asteroseismic tests are discussed. From this preliminary study, some important conclusions have already been obtained: a metallicity larger than [Fe/H] = 0.20 dex is quite unprobable and the stellar mass cannot exceed 1.22 M$_{\odot}$, contrary to Santos et al. \cite{santos04} estimate. Furthermore, models with the metallicity and the age of the Hyades are quite realistic. Asteroseismic observations of this star, which will be done in November 2006, compared with model predictions, should allow to test whether this star has been formed together with the Hyades or not. While a negative answer would be interesting in itself, a positive answer would be taken as a proof that the overmetallicity has a primordial origin.

\section{The “COROT Star” HD 52265}

For the “COROT star” HD 52265, which will be observed during five consecutive months, we expect very precise data, which will hopefully lead to very good frequency determinations and mode identifications. We will use the same kind of tests as before, and we expect to reach precise conclusions on the internal structure and past history of this exoplanets-host star. 

Preliminary computations and modelisation of HD52265 have been done using the same techniques as for $\mu$ Arae, as a preparation of the future observations with COROT (Soriano et al. \cite{soriano06}). Five different groups of observers have given external parameters for HD52265 (table2). Its luminosity is L/L$_{\odot}$=1.94$\pm$0.16. 

\begin{table}
\caption{Observed parameters for HD52265 (see Soriano et al. \cite{soriano06})a) Santos et al. 2004; b)Gonzalez et al. 2001; c) Fischer \& Valentini 2005; d) Takeda et al. 2005; e) Gillon \& Magain 2006}
\label{tab2}
\begin{center}
\begin{tabular}{cccccccc} \hline
\hline
T$_{\mbox{eff}}$ & log g & [Fe/H] & Réf. \cr
  \hline
 6103$\pm$52 & 4.28$\pm$0.12 & 0.23$\pm$0.05 & a \cr 
 6162$\pm$22 & 4.29$\pm$0.04 & 0.27$\pm$0.02 & b \cr
 6076$\pm$44 & 4.26$\pm$0.06 & 0.19$\pm$0.03 & c \cr
 6069$\pm$15 & 4.12$\pm$0.03 & 0.19$\pm$0.03 & d \cr
 6179$\pm$18 & 4.36$\pm$0.03 & 0.24$\pm$0.02 & e \cr
\hline
\end{tabular}
\end{center}
\end{table}

Evolutionary tracks have been computed for three different metallicities, as given from spectroscopic observations. Oscillation frequencies and seismic tests have been studied for specific models. Here I only show two extreme and interesting examples, for metallicities [Fe/H] = 0.19 and 0.27.

\begin{figure}
\centering
\includegraphics[width=0.8\linewidth]{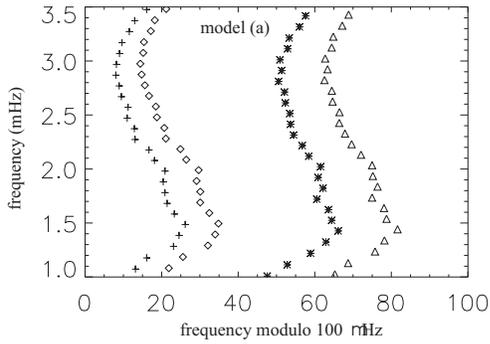}
\caption{ Echelle diagram computed for a models of HD52265 with metallicity 0.27 inside the Gonzalez et al. \cite{gonzalez01} error box (model a) (after Soriano et al. 2006). }
\label{fig3}
\end{figure}

\begin{figure}
\centering
\includegraphics[width=0.8\linewidth]{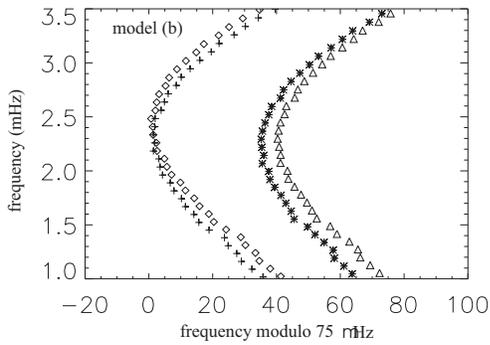}
\caption{Same as Figure 3, with metallicity 0.19 inside the Takeda et al 2005 error box (model b)  }
\label{fig4}
\end{figure}

\begin{figure}
\centering
\includegraphics[width=0.8\linewidth]{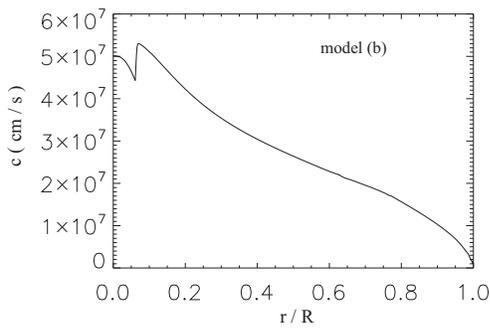}
\caption{ Profile of the sound velocity in model b, showing the special depression in the core due to the convective and helium-rich core. }
\label{fig5}
\end{figure}

\begin{figure}
\centering
\includegraphics[width=0.8\linewidth]{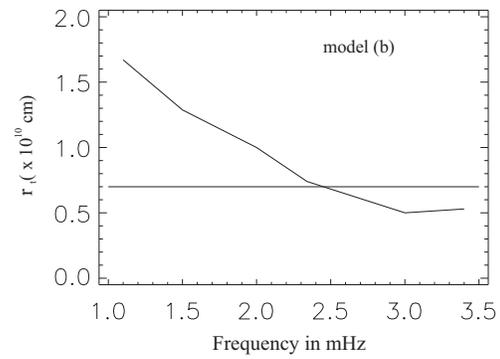}
\caption{Turning point of the $l=2$ waves as a function of the frequency: it crosses the radius of the core at the exact frequency where the crossing occurs in the echelle diagram. }
\label{fig6}
\end{figure}

The echelle diagrams for these two models are given in Figure 3 and 4. They present important differences. The most interesting feature is found in the Takeda et al. \cite{takeda05} case. The $l=0$ and $l=2$ lines cross for a frequency around 2.5 mHz, which is due to the convective, helium-rich core, as discussed in detail in Soriano et al. \cite{soriano06} (Figure 5). Indeed, the crossing point corresponds exactly to the frequency for which the turning point of the $l=2$ waves reaches the core (Figure 6). This interesting feature, which occurs only when the star is enough evolved, will be better discussed in a fothcoming paper (Vauclair \& Soriano \cite{vauclair07}).

In conclusion, this is the beginning of a new subject, asteroseismology of exoplanets-host stars, for which we expect many developments in the near future.

\end{document}